\documentclass[aps,two column,superscript address,showpacs,pre]{revtex4-1}
\usepackage{amssymb,amsfonts,bm}
\usepackage{graphics,graphicx}
\usepackage{fontenc}
\usepackage{t1enc}
\usepackage{amsmath}
\usepackage{float}
\usepackage{setspace}
\usepackage{appendix}
\usepackage{multirow}
\usepackage{color}
\usepackage{graphics,psfrag}
\usepackage{graphicx,psfrag}
\usepackage{enumerate}
\usepackage[none]{hyphenat}

\DeclareMathOperator{\Tr}{Tr}

\begin{document}

\title{The original electric-vertex formulation of the symmetric eight-vertex 
model on the square lattice is fully non-universal}

\author{Roman Kr\v{c}m\'ar}
\author{Ladislav \v{S}amaj}
\affiliation{Institute of Physics, Slovak Academy of Sciences, 
D\'ubravsk\'a cesta 9, SK-84511, Bratislava, Slovakia}

\begin{abstract}
The partition function of the symmetric (zero electric field) eight-vertex 
model on a square lattice can be formulated either in the original 
``electric'' vertex format or in an equivalent ``magnetic'' Ising-spin format.
In this paper, both electric and magnetic versions of the model are studied 
numerically by using the Corner Transfer Matrix Renormalization Group method 
which provides reliable data.
The emphasis is put on the calculation of four specific critical exponents,
related by two scaling relations, and of the central charge.
The numerical method is first tested in the magnetic format, the obtained 
dependences of critical exponents on model's parameters agree with 
Baxter's exact solution and weak universality is confirmed within 
the accuracy of the method due to the finite size of the system. 
In particular, the critical exponents $\eta$ and $\delta$ are constant
as required by weak universality.
On the other hand, in the electric format, analytic formulas based on the
scaling relations are derived for the critical exponents $\eta_{\rm e}$ and 
$\delta_{\rm e}$ which agree with our numerical data. 
These exponents depend on model's parameters which is an evidence for the full 
non-universality of the symmetric eight-vertex model in the original 
electric formulation.
\end{abstract}


\maketitle

\renewcommand{\theequation}{1.\arabic{equation}}
\setcounter{equation}{0}

\section{Introduction} \label{Sect.1}
The two-dimensional (2D) eight-vertex model on the square lattice was proposed 
as a generalization of ice-type systems in 1970 \cite{Sutherland70,Fan70}.
Its symmetric (zero electric field) version was solved by using the idea of 
commuting transfer matrices and the Yang-Baxter equation for the scattering 
matrix as the consistency condition 
\cite{Baxter71,Baxter72a,Baxter72b,Baxterbook}. 
This became a basis for generating and solving systematically integrable 
models within the so-called ``Quantum Inverse-Scattering method'' (QISM)
\cite{Sklyanin78a,Sklyanin78b}, see monographs \cite{Korepinbook,Samajbook}. 

The partition function of the original ``electric'' eight-vertex formulation
can be mapped onto the partition function of a ``magnetic'' Ising model on 
the dual square lattice with plaquette interactions \cite{Wu71,Kadanoff71}.  
The exact magnetic critical exponents of the symmetric eight-vertex model 
depend continuously on model's parameters \cite{Baxterbook}.
This violates the universality hypothesis which states that critical 
exponents of a statistical system depend only on the symmetry of microscopic 
state variables and the spatial dimensionality of the system \cite{Griffiths70}.
Suzuki \cite{Suzuki74} formulated the singularities of statistical 
quantities near the critical point not in terms of the usual temperature 
difference, but in terms of the inverse correlation length which also goes 
to zero when approaching the critical point.
The rescaled critical exponents are universal; this phenomenon is known
as ``weak universality''.
The necessary condition for weak universality is the constant value of 
critical exponents defined just at the critical point, namely $\eta$ and 
$\delta$, since the freedom in the definition of deviation from the critical 
point has no effect on these exponents. 

Kadanoff and Wegner \cite{Kadanoff71} suggested that the variation of
critical indices is due to the special hidden symmetries of the zero-field 
eight-vertex model. 
If an external field is applied, they argued that the magnetic exponents 
should be constant and equivalent to those of the standard 2D Ising model,
see also monograph \cite{Baxterbook}.
This conjecture was supported by renormalization group calculations
\cite{Leeuwen75,Kadanoff79,Knops80}.
Recently, the conjecture was confirmed numerically, except for
two specific ``semi-symmetric'' combinations of vertical and horizontal 
electric fields for which the model still exhibits weak universality
\cite{Krcmar16}. 

Historically, the next weakly universal Ashkin-Teller model 
\cite{Ashkin43,Fan72,Kadanoff77,Zisook80} is in fact related to 
the eight-vertex model \cite{Kadanoff79}.
Weak universality appears also in interacting dimers \cite{Alet05}, 
frustrated spins \cite{Queiroz11,Jin12}, quantum phase transitions 
\cite{Suzuki15}, models of percolation \cite{Andrade13}, etc.
There are indications that both universality and weak universality
are violated in the symmetric 16-vertex model on the 2D square and 
three-dimensional (3D) diamond lattices \cite{Kolesik93a,Kolesik93b}, 
Ising spin glasses \cite{Bernardi95}, frustrated spin models 
\cite{Bekhechi03}, experimental measurements on composite materials 
\cite{Omerzu01,Kagawa05}, etc.  

The six-vertex model is a simplified ice-type version of the eight-vertex
model with certain vertex weights equal to zero.
This model, represented as the quantum Heisenberg XXZ spin-$\frac{1}{2}$ chain,
is related to many other systems like supersymmetric spin chains
\cite{Kulish80,Chalker88,Lee94,Kondev97,Fendley03,Hagendorf12},
2D loop and tilling models \cite{Gier04,Nichols05,Zinn-Justin10}, 
the random-cluster model of Fortuin and Kasteleyn
\cite{Fortuin72a,Fortuin72b,Grimmett06}, the restricted solid-on-solid
model \cite{Pasquier87,Owczarek87} and classical 2D Potts models
\cite{Wu82,Jacobsen06}.
The relations of the six-vertex model to these models have a precise
meaning within Temperley-Lieb algebra representation theory \cite{Temperley71}.
Although all partition functions of the related models are equal, 
the content of critical exponents is only partially overlapping.

The polarization is an order parameter in the symmetric eight-vertex model. 
The corresponding critical exponent $\beta_{\rm e}$, which depends on model's
parameters, is the only exactly known electric exponent \cite{Baxterbook}. 
The restriction to the six-vertex model and the related XXZ spin chain 
provides an additional information about electric critical exponents. 
Using previous results about the arrow correlation length exponent for the
six-vertex model \cite{Johnson73}, Luther and Peschel \cite{Luther75} have 
shown that the arrow correlation function is the same as the transverse spin 
correlation of the Heisenberg XXZ model.
Using a generalization of the Jordan-Wigner transformation for spin operators,
they were able to calculate the asymptotic behavior of spin correlation
functions for a continuum generalization of the spin-$\frac{1}{2}$ XXZ chain
and suggested a formula for indices $\gamma_{\rm e}$ and $\eta_{\rm e}$.  
The analytical predictions for the electric critical indices was verified 
well numerically by using the Trotter approximation \cite{Takada86}.
The only numerical complication concerns the isotropic XXX antiferromagnetic
chain where a multiplicative logarithmic correction for the correlation
function exists; for a controversial discussion about this topic see Refs. 
\cite{Kubo88,Liang90,Lin91}.
A density-matrix renormalization-group study \cite{Hallberg95} improved
the previous calculations of the logarithmic correction. 
 
To our knowledge, no direct numerical studies of the electric critical
exponents have been made for the eight-vertex model.  
The aim of the present paper is to study numerically both magnetic and electric
critical exponents of the symmetric eight-vertex model.
To achieve a high accuracy, we apply the Corner Transfer Matrix 
Renormalization Group (CTMRG) method, based on the renormalization of 
the density matrix \cite{White92,White93,Schollwock05,Krcmar15}.
Four critical exponents which fulfill two scaling relations and the central
charge are calculated in both magnetic and electric formats. 
The CTMRG method is first tested on the magnetic version of the symmetric 
eight-vertex model, the obtained dependence of magnetic critical exponents 
on model's parameters is in good agreement with Baxter's exact solution 
and weak universality is verified.
In particular, the critical exponents $\eta$ and $\delta$ are constant, 
as required by weak universality.
On the other hand, in the electric format, analytic formulas based on the
scaling relations are derived for the critical exponents $\eta_{\rm e}$ and 
$\delta_{\rm e}$ which agree with our numerical data. 
These exponent depends on model's parameters which is an evidence that 
both universality and weak universality are violated, i.e.,
the original electric formulation of the eight-vertex model is fully 
non-universal.
Thus the equivalence of the electric and magnetic partition functions 
does not imply the same critical properties of the two model versions.

The paper is organized as follows.
In Sec. \ref{Sec.2}, we summarize basic facts about the symmetric
eight-vertex model on the square lattice.
These facts include the mapping onto the Ising model with plaquette
interactions, definitions of critical exponents of interest and of 
their scaling relations and the exact results of Baxter. 
In Sec. \ref{Sec.3}, we review briefly the CTMRG numerical method and 
the evaluation techniques of magnetic and electric critical exponents.
The numerical method is first tested on magnetic critical exponents
in Sec. \ref{Sec.4}, their dependences on model's parameters agree with 
Baxter's values and phenomenon of weak universality is checked within 
the accuracy of the method due to the finite size of the system. 
The numerical results for electric counterparts of critical exponents,
presented in Sec. \ref{Sec.5}, confirm clearly that the symmetric eight-vertex
model is fully non-universal in its original vertex format.
Sec. \ref{Sec.6} brings a brief recapitulation.

\renewcommand{\theequation}{2.\arabic{equation}} 
\setcounter{equation}{0}

\section{Basic facts about the symmetric eight-vertex model} \label{Sec.2}
\begin{figure}
\centering
\includegraphics[width=0.45\textwidth,clip]{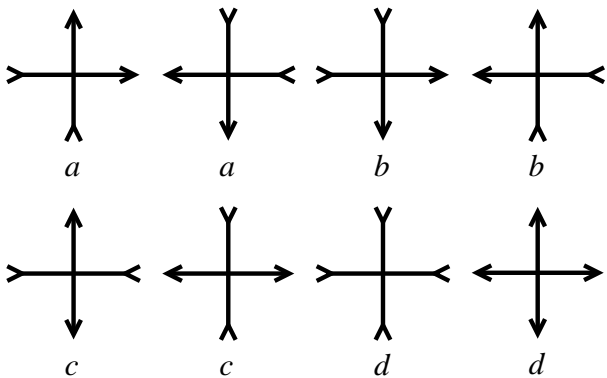}
\caption{Admissible configurations of the eight-vertex model, with the
corresponding notation of the Boltzmann vertex weight.}
\label{obr:vertexy1}
\end{figure}

In vertex models, one attaches to each lattice edge local two-state variables,
say arrows directing to the one of two vertices joint by the edge;
the arrows can be interpreted as electric dipoles. 
In the eight-vertex model, each vertex configuration of edge states
satisfies the rule that only even (0, 2 or 4) number of arrows point toward 
the vertex.
From among $2^4=16$ possible vertex configurations just eight ones fulfill 
this rule; the admissible configurations of arrows together with the 
corresponding Boltzmann vertex weights are presented in Fig. \ref{obr:vertexy1}.
In the symmetric version of the eight-vertex model considered here, 
the Boltzmann weight of a vertex configuration is invariant with 
respect to the reversal of all arrows incident to a vertex which corresponds
to zero electric fields acting on dipole arrows.
The Boltzmann vertex weights can be formally expressed in terms of local 
energies as follows
\begin{equation} \label{fieldrepr}
\begin{split}
a = \, \, C \exp\left( - \epsilon_a/T \right) , \qquad
b = \, \, C \exp\left( - \epsilon_b/T \right) , \\
c = \, \, C \exp\left( - \epsilon_c/T \right) , \qquad
d = \, \, C \exp\left( - \epsilon_d/T \right) ,
\end{split}
\end{equation}
where $T$ is the temperature (in units of $k_{\rm B}=1$) and the value of 
the prefactor $C$ is irrelevant.
The partition function is defined by
\begin{equation} \label{part}
Z_{\rm 8V}(T) = \sum \prod \mbox{(weights)} ,
\end{equation}
where the summation goes over all possible edge configurations on the lattice
and, for a given configuration, the product is taken over all vertex weights.

\subsection{Mapping onto the Ising model}
\begin{figure}
\centering
\includegraphics[width=0.25\textwidth,clip]{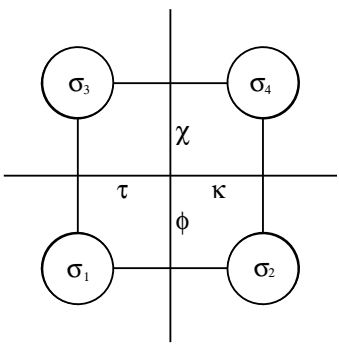}
\caption{Mapping from the electric vertex formulation with edge states
$\phi,\chi,\tau,\kappa$ to the magnetic Ising representation
with site spin variables $\sigma_1,\sigma_2,\sigma_3,\sigma_4$.}
\label{obr:prepis}
\end{figure}

The symmetric eight-vertex model on the square lattice can be mapped onto its 
Ising counterpart defined on the dual (also square) lattice 
\cite{Wu71,Kadanoff71}.
We assign $+1$ to the up/right arrows and $-1$ to the down/left arrows.
A state configuration $\phi,\chi,\tau,\kappa$ ($\phi=\pm 1$,
$\chi=\pm 1$, etc.) of incident edges is depicted in Fig. \ref{obr:prepis}.
The eight-vertex rule is equivalent to the constraint
\begin{equation} \label{req}
\phi \chi \tau \kappa = 1 .
\end{equation}
The Ising spin variables on the dual square
$\sigma_1,\sigma_2,\sigma_3,\sigma_4$ ($\sigma_1=\pm 1$, $\sigma_2=\pm 1$, etc.)
are related to the vertex edge variables at the bond intersections as follows
\begin{equation} \label{mapping}
\phi = \sigma_1\sigma_2 , \quad \chi = \sigma_3\sigma_4 , \quad
\tau = \sigma_1\sigma_3 , \quad \kappa = \sigma_2\sigma_4 .
\end{equation}
Due to the equality 
$\phi\chi\tau\kappa = \sigma_1^2\sigma_2^2\sigma_3^2\sigma_4^2$
the eight-vertex requirement (\ref{req}) is automatically fulfilled.
Note that the spin-flip transformation $\sigma_i\to -\sigma_i$ for all
$i=1,2,3,4$ leaves the actual values of vertex states unchanged.

The Ising Hamiltonian can be written as 
\begin{equation}
H_{\rm I} = \sum_{\rm plaq} H_{\rm plaq} ,
\end{equation}
where each square plaquette Hamiltonian $H_{\rm plaq}$ involves interactions
of four spins $\sigma_1,\sigma_2,\sigma_3,\sigma_4=\pm 1$ as depicted in 
Fig. \ref{obr:prepis}. 
The plaquette Hamiltonian involves diagonal and four-spin interactions, 
\begin{equation} \label{eq:ising_ham}
- H_{\rm plaq} = J\sigma_2\sigma_3  + J'\sigma_1\sigma_4
+ J'' \sigma_1\sigma_2\sigma_3\sigma_4 . 
\end{equation}
It exhibits the spin-flip symmetry $\sigma_i\to -\sigma_i$ ($i=1,2,3,4$).

The partition function of the eight-vertex model (\ref{part}) and the one of
the Ising model 
\begin{equation}
Z_{\rm I}(T) = \sum_{\{\sigma\}} \exp(-H_{\rm I}/T)
\end{equation}
are equivalent,
\begin{equation} \label{partitioneq}
Z_{\rm I}(T) = 2 Z_{\rm 8V}(T) ,
\end{equation}
if the Boltzmann vertex weights are expressed in terms of 
the Ising interactions in the following way \cite{Baxterbook} 
\begin{equation} \label{eq:magneticrepre1}
\begin{split}
a = & \, \, C \exp\left[ \left( J+J'+J''\right)/T \right] , \\
b = & \, \, C \exp\left[ \left( -J-J'+J''\right)/T \right] , \\
c = & \, \, C \exp\left[ \left( -J+J'-J'' \right)/T \right] , \\
d = & \, \, C \exp\left[ \left( J-J'-J'' \right)/T \right] .
\end{split}
\end{equation}
The Boltzmann vertex weight $w(\tau,\phi\vert \chi,\kappa)$, corresponding
to the configuration of edge state in Fig. \ref{obr:prepis}, which are 
constrained by (\ref{req}), is expressible in terms of Ising couplings
as follows 
\begin{equation} \label{w}
w(\tau,\phi\vert \chi,\kappa) = \exp\left[ \left(
J \phi\tau + J' \chi\tau + J''\phi\chi \right)/T \right] .
\end{equation}

In terms of the free energy $F$ defined as $-F/T = \ln Z$, the relation 
between the partition functions (\ref{partitioneq}) is equivalent to
\begin{equation}
- F_{\rm I}(T)/T = \ln 2 - F_{\rm 8V}(T)/T . 
\end{equation}
For the internal energies defined by $U = -T^2\partial (F/T)\partial T$,
it holds that
\begin{equation} \label{internal}
U_{\rm I}(T) = U_{\rm 8V}(T) . 
\end{equation}

Since the Ising Hamiltonian $H_{\rm I}$ is invariant with respect to 
the spin-flip transformation $\sigma_i\to -\sigma_i$ at all lattice sites, 
the Ising magnetization
\begin{equation}
M = \langle \sigma \rangle
\end{equation} 
($\langle\cdots\rangle$ means the thermodynamic average) 
is a good order parameter in the ferromagnetic phase.

For every state configuration of edges incident to each vertex, the constraints 
(\ref{req}) and the Boltzmann weights (\ref{w}) are invariant with respect
to the transformation $\phi,\chi,\tau,\kappa\to -\phi,-\chi,-\tau,-\kappa$.
The isotropic polarization
\begin{equation} \label{polar}
P = \langle \phi \rangle .
\end{equation}
is therefore a legitimate order parameter as well.
Note that due to the relations between the arrow and spin variables 
(\ref{mapping}), the polarization is equal to the correlation function 
of nearest-neighbor Ising spins.

\subsection{Magnetic format: exact results}
The symmetric eight-vertex model has five phases \cite{Baxterbook}.
We shall restrict ourselves to the ferroelectric-A phase defined by 
the inequality $a > b + c + d$ and the disordered phase in the region
$a,b,c,d < (a + b + c + d)/2$. 
The second-order transition between these phases takes place at 
the hypersurface 
\begin{equation} \label{crit}
a_c=b_c+c_c+d_c ,
\end{equation}
where $c$-subscript means evaluated at the critical temperature $T_c$.
Note that our vertex weights do not belong to the ``principal regime''
defined by the inequality $c>a+b+d$ (see Sec. 10.7 of monograph
\cite{Baxterbook}), so certain formulas in \cite{Baxterbook} written
for vertex weights in the principal regime must be adapted to our case.  

In general, only two critical exponents are independent and all other 
exponents can be expressed in terms of them by using scaling relations 
\cite{Baxterbook}.
Here, we shall concentrate on four critical exponents.

Let us consider a small temperature deviation from the critical point 
$\Delta T = T-T_c$. 
For $\Delta T\to 0^-$, the spontaneous magnetization $M$ behaves as
\begin{equation} \label{sponmagn}
M \propto (-\Delta T)^{\beta}
\end{equation} 
which defines the critical index $\beta$.

The pair spin-spin correlation function at distance $r$, 
$G({\bf r}) = \langle \sigma_{\bf 0}\sigma_{\bf r} \rangle$, has in 2D 
the large-distance asymptotic form
\begin{equation} \label{G}
G(r) \propto \frac{1}{r^{\eta}} \exp\left( -r/\xi \right) ,
\end{equation}  
where $\xi$ is the correlation length.
Approaching the critical point, the correlation length diverges as
\begin{equation}
\xi \mathop{\propto}_{\Delta T\to 0^+} \frac{1}{(\Delta T)^{\nu}} , \qquad
\xi \mathop{\propto}_{\Delta T\to 0^-} \frac{1}{(-\Delta T)^{\nu'}} ,
\end{equation}
where the critical exponents $\nu$ and $\nu'$ are in fact identical.
Just at the critical point, where $\xi\to\infty$, the exponential short-range 
decay of the correlation function (\ref{G}) becomes long-ranged,
\begin{equation} \label{Gcrit}
G(r) \propto \frac{1}{r^{\eta}} , \qquad T=T_c
\end{equation}  
which defines the exponent $\eta$.

Let us apply to the spin system an external magnetic field $H$, so that
the Ising Hamiltonian can be written as
\begin{equation}
H_{\rm I} = \sum_{\rm plaq} H_{\rm plaq} - H \sum_i \sigma_i .
\end{equation}
The critical point corresponds to $T=T_c$ and $H=0$.
At $T=T_c$ and for small $H$, the Ising magnetization $M(H)$ exhibits
the singular behavior of type
\begin{equation} \label{Hcrit}
M(H) \propto H^{1/\delta} , \qquad T = T_c
\end{equation}
which defines the critical exponent $\delta$.

The von Neumann entropy is defined by
\begin{equation} \label{Neumann}
S_{\rm N} = - {\rm Tr}\, \rho \ln \rho ,
\end{equation}
where $\rho$ is the density matrix of the Ising model defined below.
At the critical point, the entropy grows with the size $L$ of the system as
\cite{Calabrese04,Ercolessi10}
\begin{equation}
S_{\rm N} \sim \frac{c}{6} \ln L , \qquad T=T_c ,
\end{equation}
where $c$ is the central charge.
It holds that $c=1$ for the weakly universal symmetric eight-vertex model
\cite{Baxterbook}.
We recall that $c=1/2$ for the universal 2D Ising model. 

In 2D, the four exponents of interest fulfill two scaling relations 
\cite{Baxterbook} 
\begin{equation} \label{scaling}
\eta = 2 \frac{\beta}{\nu} , \qquad \delta = \frac{4}{\eta} - 1 .
\end{equation}

According to the exact Baxter's solution of the symmetric eight-vertex model,
the exponents $\beta$ and $\nu$, whose definition requires to introduce 
the small temperature deviation $\Delta T$, are given by \cite{Baxterbook}
\begin{equation} \label{eq:crit_ex}
\beta = \frac{\pi}{16\mu} , \qquad
\nu = \frac{\pi}{2\mu} , 
\end{equation}
where the auxiliary parameter
\begin{equation} \label{eq:mu}
\mu = 2 \arctan \left( \sqrt{\frac{a_cb_c}{c_cd_c}} \right)
= 2 \arctan \left( e^{2J''/T_c} \right) .
\end{equation}
If $J''=0$, when the system splits into two independent Ising lattices
with nearest-neighbor couplings $J$ and $J'$, we have $\mu=\pi/2$ and 
Eq. (\ref{eq:crit_ex}) gives the standard 2D Ising exponents 
\begin{equation} \label{eq:crit_ex1}
\beta_{\rm I} = \frac{1}{8} , \qquad \nu_{\rm I} = 1 . 
\end{equation}

Suzuki's concept of weak universality \cite{Suzuki74} explains the dependence 
of the critical exponents (\ref{eq:crit_ex}) on $J''$ by the ambiguity in 
the definition of the deviation from the critical point.
If one considers the inverse correlation length $\xi^{-1}\propto (T_c-T)^{\nu}$
with $T\to T_c^-$ instead of the temperature difference $T_c - T$,
the new (rescaled) critical exponent 
\begin{equation} \label{eq:crit_del}
\hat{\beta} \equiv \frac{\beta}{\nu} = \frac{1}{8}
\end{equation}
becomes universal.
According to the definitions (\ref{Gcrit}) and (\ref{Hcrit}), the exponents 
$\eta$ and $\delta$ are defined just at the critical point and as such do not 
depend on the definition of the deviation from the critical point.
Therefore $\eta$ and $\delta$ must be constant in a weakly universal theory
and this fact is confirmed by Baxter's result 
\begin{equation} \label{eta}
\eta = \frac{1}{4} , \qquad \delta = 15 ,
\end{equation}
i.e., $\eta=\eta_{\rm I}$ and $\delta=\delta_{\rm I}$.
The scaling relations (\ref{scaling}) evidently holds
for the exponents (\ref{eq:crit_ex}) and (\ref{eta}).

\subsection{Electric format: exact results}
As concerns the electric format, the only exactly known critical exponent
\cite{Baxterbook}
\begin{equation} \label{betae}
\beta_{\rm e} = \frac{\pi-\mu}{4\mu} ,
\end{equation}
with $\mu$ defined by Eq. (\ref{eq:mu}), describes the singular behavior of 
the spontaneous polarization near the critical point,
\begin{equation}
P \propto (-\Delta T)^{\beta_e} .
\end{equation}
In order to distinguish between magnetic and electric exponents, we add 
the subscript ``e'' to the latter.

In analogy with the magnetic system, we introduce the pair arrow-arrow 
correlation function at distance $r$, 
$G_{\rm e}({\bf r}) = \langle \phi_{\bf 0}\phi_{\bf r} \rangle$. 
In 2D, it exhibits the large-distance behavior of type  
\begin{equation} \label{Ge}
G_{\rm e}(r) \propto \frac{1}{r^{\eta_{\rm e}}} \exp\left( -r/\xi_{\rm e} \right) .
\end{equation}  
Close to the critical point, the correlation length $\xi_{\rm e}$ diverges as
\begin{equation}
\xi_{\rm e} \mathop{\propto}_{\Delta T\to 0^+} \frac{1}{(\Delta T)^{\nu_{\rm e}}} , 
\qquad
\xi_{\rm e} \mathop{\propto}_{\Delta T\to 0^-} \frac{1}{(-\Delta T)^{\nu'_{\rm e}}} , 
\end{equation}
where $\nu_{\rm e}=\nu'_{\rm e}$. 
At the critical point, 
\begin{equation}
G(r) \propto \frac{1}{r^{\eta_{\rm e}}} , \qquad T=T_c .
\end{equation}

Let us apply an isotropic electric field $E_x=E_y=E$, so that the Hamiltonian 
changes by $-E\sum_{\langle i,j\rangle} \phi_{\langle i,j\rangle}$ where 
$\phi_{\langle i,j\rangle}$ is the state variable on the edge connecting 
nearest-neighbor sites $i$ and $j$. 
The critical point corresponds to $T=T_c$ and $E=0$.
At $T=T_c$ and for small $E$, the polarization $P(E)$ behaves as
\begin{equation} \label{Ecrit}
P(E) \propto E^{1/\delta_{\rm e}} , \qquad T = T_c
\end{equation}
which defines the electric exponent $\delta_{\rm e}$.

As in the magnetic format, the von Neumann entropy $S^{\rm e}_{\rm N}$
is defined by (\ref{Neumann}) with $\rho$ being the density matrix
of the vertex model.
At the critical point, the entropy grows with the system size $L$ as
\begin{equation}
S^{\rm e}_{\rm N} \sim \frac{c_{\rm e}}{6} \ln L , \qquad T=T_c ,
\end{equation}
where $c_{\rm e}$ is the electric central charge.

The four electric critical exponents of interest fulfill the electric 
counterparts of scaling relations (\ref{scaling}):
\begin{equation} \label{scalinge}
\eta_{\rm e} = 2 \frac{\beta_{\rm e}}{\nu_{\rm e}} , \qquad
\delta_{\rm e} = \frac{4}{\eta_{\rm e}} - 1 .
\end{equation}

\renewcommand{\theequation}{3.\arabic{equation}}
\setcounter{equation}{0}

\section{Numerical method} \label{Sec.3}
\subsection{CTMRG approach}
\begin{figure}
\centering
\includegraphics[width=0.4\textwidth,clip]{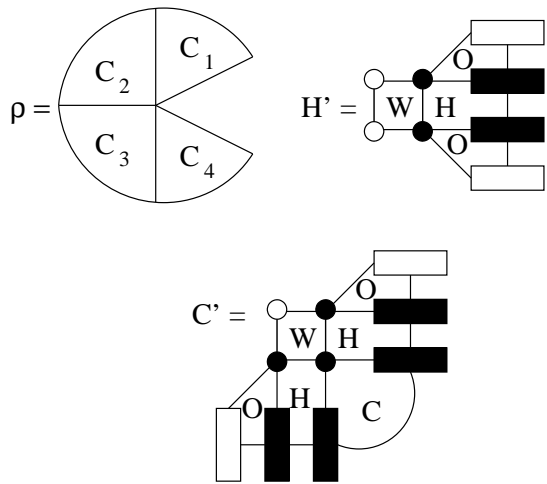}
\caption{The CTMRG renormalization process. 
The density matrix $\rho$ is composed of four transfer matrices $C_1$, $C_2$,
$C_3$ and $C_4$; each straight line represents $L$ matrix site indices which 
are either fixed (``free'' lines adjacent to $C_1$ and $C_4$) or summed out 
[common lines of pairs $(C_1,C_2)$, $(C_2,C_3)$ and $(C_3,C_4)$]. 
The expansion process of the corner transfer matrix $C$ and the half-row 
transfer matrix $H$ from the previous iteration, see text.}
\label{obr:ctmrg}
\end{figure}

The CTMRG method \cite{Nishino96a,Nishino97} is based on Baxter's technique
of corner transfer matrices \cite{Baxterbook}. 
Each quadrant of the square lattice with size $L\times L$ is represented 
by one of the corner transfer matrices $C_1,\cdots,C_4$ and 
the partition function $Z = \Tr (C_1C_2C_3C_4)$. 
The density matrix is defined by $\rho = C_1C_2C_3C_4$, 
see Fig. \ref{obr:ctmrg}, so that $Z = \Tr\rho$. 
The number of degrees of freedom grows exponentially with $L$ and 
the density matrix is used in the process of their reduction. 
Namely, degrees of freedom are iteratively projected to the space generated 
by the eigenvectors of the density matrix with the largest eigenvalues. 
The projector on this reduced space of dimension $D$ will be denoted by $O$; 
the larger truncation parameter $D$ is taken, the better precision of 
the results is attained. 
In each iteration the linear size of the system is expanded from $L$ to $L+2$
via the inclusion of the Boltzmann weight $W$ of the basic plaquette cell 
(see Fig. \ref{obr:prepis}).
The expansion process transforms the corner transfer matrix $C$ to $C'$
and the half-row transfer matrix $H$ to $H'$ in the way represented 
schematically in Fig. \ref{obr:ctmrg}.
The empty boxes (circles) represent new multi-spin (spin) variables 
obtained after the renormalization which consists in the summation and 
$O$-projection of multi-spin (spin) black boxes (circles) from 
the previous iteration. 
The fixed boundary conditions are used, with each spin at the boundary set  
to the value $\sigma = -1$. 
This choice ensures a quicker convergence of the method in the ordered phase.

Technically, one has to distinguish between two choices of
vertex weights $c$ and $d$.

The choice
\begin{equation} \label{cd}
c = d
\end{equation}
leads to the symmetric density matrix $\rho$.
In the Ising representation (\ref{eq:magneticrepre1}), 
the choice (\ref{cd}) corresponds to the constraint $J=J'$.
The original formulation of CTMRG \cite{Nishino96a,Nishino97} requires
that the density matrix $\rho$ is symmetric. 
In that case we can return to the row-to-row transfer matrix ${\cal T}$ and 
denote by $\vert\psi_0^{(l)}\rangle$ and $\vert\psi_0^{(r)}\rangle$ its left and
right eigenvectors corresponding to the largest eigenvalue, respectively.
For the symmetric ${\cal T}$, we have the equality
$\vert\psi_0^{(l)}\rangle = \vert\psi_0^{(r)}\rangle$.  
In the limit $L\to\infty$, the product of the corner matrices $C_1 C_2$ 
is expressible as $\vert\psi_0^{(l)}\rangle$ and $C_3 C_4$ as 
$\langle\psi_0^{(r)}\vert$, so that
\begin{equation}
\rho = \Tr \vert\psi_0^{(l)}\rangle \langle\psi_0^{(r)}\vert .
\end{equation}  
Here, the trace is taken over common indices of the corner matrices 
$C_2$ and $C_3$, see Fig. \ref{obr:ctmrg}.

If
\begin{equation} \label{cnd}
c \neq d ,
\end{equation}
it holds that $\vert\psi_0^{(l)}\rangle \neq \vert\psi_0^{(r)}\rangle$
and the density matrix is non-symmetric.
Within the Ising representation (\ref{eq:magneticrepre1}), 
this choice of vertex weights corresponds to the inequality $J\neq J'$. 
It can be shown \cite{White93,Carlon99} that the symmetrized density matrix
\begin{equation} \label{symrho}
\rho = \frac{1}{2} \Tr \left(
\vert\psi_0^{(l)}\rangle \langle\psi_0^{(l)}\vert +
\vert\psi_0^{(r)}\rangle \langle\psi_0^{(r)}\vert \right)
\end{equation}
provides an optimal basis set which minimizes the distance of a trial
vector in the reduced space (of dimension $D$) from the right and left
eigenstates $\vert\psi_0^{(l)}\rangle$ and $\vert\psi_0^{(r)}\rangle$. 
This fact allows us to use the symmetrized density matrix (\ref{symrho}) 
within the standard CTMRG \cite{Nishino96a,Nishino97} when treating 
the more complicated case (\ref{cnd}).
The only exception is the von Neumann entropy (\ref{Neumann}) for which
the above approach does not work and therefore in this case we shall consider 
only the choice $c=d$ with the symmetric density matrix $\rho$.

\subsection{Calculation of critical exponents}
First we focus on the magnetic critical exponents $\nu$, $\eta$, $\beta$,
$\delta$ and the central charge $c$, and then on their electric counterparts.

\subsubsection{Magnetic exponents}
The critical magnetic exponent $\nu$ can be obtained from the dependence of 
the internal energy $U_{\rm I}$ on the linear size of the system $L$ 
at the critical point \cite{Nishino96b},
\begin{equation} \label{UL}
U_{\rm I}(L) - U_{\rm I}(\infty) \propto L^{-2+1/\nu} , \qquad T = T_c .
\end{equation}
The effective (i.e., $L$-dependent) exponent $\nu^{\rm eff}$ is calculated 
as the logarithmic derivative of the internal energy as follows
\begin{equation}\label{eq:nu0f}
\nu^{\rm eff}(L) = \left[ 3 + \frac{\partial}{\partial\ln L}
\ln\left(\frac{\partial U_{\rm I}}{\partial L}\right) \right]^{-1} .
\end{equation}
If $T$ is close to the critical $T_c$, the plot $\nu^{\rm eff}(L)$ either goes 
to 0 (in the ordered phase) or diverges (in the disordered phase) with 
increasing $L$.
We can therefore determine the critical temperature $T_c$ from
the requirement that $\nu^{\rm eff}(L)$ goes to a finite non-zero value
as $L\to\infty$, i.e.,
\begin{equation} \label{stab}
\lim_{L\to\infty} \nu^{\rm eff}(L) \to \nu , \qquad T=T_c ,
\end{equation}
where $0<\nu<\infty$ is the critical exponent we are searching for. 
For the model under consideration with the known critical manifold
this procedure is not necessary, but we checked that it reproduces with 
a high precision the exact relation (\ref{crit}).
 
The magnetic index $\eta$ follows from the $L$-dependence of the magnetization 
at the critical point \cite{Nishino96b}, 
\begin{equation} \label{eq:fss}
M \mathop{\propto}_{L\to\infty}  L^{-\eta/2} , \qquad T = T_c .
\end{equation}
The effective exponent $\eta^{\rm eff}$ is calculated as the logarithmic 
derivative of magnetization
\begin{equation} \label{eq:eta0f}
\eta^{\rm eff}(L) = -2\frac{\partial \ln M}{\partial \ln L} .
\end{equation}
As before, $\eta = \lim_{L\to\infty} \eta^{\rm eff}(L)$.

To calculate the magnetic exponent $\beta$, we make use of the $T$-dependence 
of the spontaneous magnetization $M$ close to the critical temperature $T_c$,
see Eq. (\ref{sponmagn}).
The effective exponent $\beta^{\rm eff}$ is extracted via the logarithmic 
derivative
\begin{equation} \label{eq:beta0f}
\beta^{\rm eff}(T) = \frac{\partial\ln M}{\partial\ln(T_c-T)} .
\end{equation}
In general, $\beta^{\rm eff}$ as a function of $T$ has one extreme (maximum) 
at $T^*$, decays slowly for $T<T^*$ and drops abruptly for $T^*<T<T_c$, 
as a sign that the CTMRG method is inaccurate close to $T_c$. 
The extreme condition $\partial \beta^{\rm eff}/\partial T\vert_{T=T^*} = 0$
indicates a weak dependence of $\beta^{\rm eff}$ on $T$ close to $T^*$.
This is why we take as the critical index $\beta$ the maximal value of 
$\beta^{\rm eff}$, $\beta = \beta^{\rm eff}(T^*)$.

To obtain the magnetic exponent $\delta$, we recall that 
the magnetization $M$ behaves at the critical temperature $T=T_c$ 
according to the relation (\ref{Hcrit}).
The effective exponent $\delta^{\rm eff}$ is calculated as follows
\begin{equation} \label{eq:delta0f}
\delta^{\rm eff}(H) = \left( \frac{\partial\ln M}{\partial\ln H} \right)^{-1}.
\end{equation}
In close analogy with the case of $\beta^{\rm eff}$, $\delta^{\rm eff}$
as a function of $H$ has one extreme (minimum) at $H^*$ and
$\delta = \delta^{\rm eff}(H^*)$.

As concerns the von Nemann entropy (\ref{Neumann}), at $T=T_c$ we define
the effective central charge
\begin{equation}
c^{\rm eff}(L) = 6 \frac{\partial S_{\rm N}}{\partial \ln L}
\end{equation}
and $c=\lim_{L\to\infty} c^{\rm eff}(L)$.

\subsubsection{Electric exponents}
Now we pass to the electric critical exponents.
The critical index $\nu_{\rm e}$ can be calculated from the electric
counterpart of Eq. (\ref{UL})
\begin{equation} \label{ULe}
U_{\rm 8V}(L) - U_{\rm 8V}(\infty) \propto L^{-2+1/\nu_{\rm e}} , \qquad T = T_c .
\end{equation}
Choosing the equivalent boundary conditions, the relation between 
the Ising and vertex internal energies (\ref{internal}) can be adopted 
for any system size $L$, 
\begin{equation} \label{internalL}
U_{\rm I}(T,L) = U_{\rm 8V}(T,L) . 
\end{equation}
In view of relations (\ref{UL}) and (\ref{ULe}), the corresponding magnetic 
and electric exponents coincide:
\begin{equation} \label{nue}
\nu_{\rm e} = \nu .
\end{equation}

The critical electric index $\eta_{\rm e}$ follows from the large-$L$
dependence of the polarization at the critical point \cite{Nishino96b}, 
\begin{equation} \label{eq:fssp}
P \propto  L^{-\eta_{\rm e}/2} , \qquad T = T_c .
\end{equation}
The effective exponent $\eta_{\rm e}^{\rm eff}$ is calculated as 
\begin{equation} \label{eq:eta0fp}
\eta_{\rm e}^{\rm eff}(L) = -2\frac{\partial \ln P}{\partial \ln L} .
\end{equation}
Finally, $\eta_{\rm e} = \lim_{L\to\infty} \eta_{\rm e}^{\rm eff}(L)$.

Taking into account that below the critical temperature the spontaneous 
polarization $P$ behaves as
\begin{equation} \label{eq:magnetp}
P \propto (T_c-T)^{\beta_{\rm e}} \qquad \mbox{as $T\to T_c^-$,}
\end{equation}
the effective exponent $\beta_{\rm e}^{\rm eff}$ is retrieved via 
\begin{equation} \label{eq:beta0fp}
\beta_{\rm e}^{\rm eff}(T) = \frac{\partial\ln P}{\partial\ln(T_c-T)} .
\end{equation}
The critical index $\beta_{\rm e}$ corresponds to the maximal value of 
$\beta_{\rm e}^{\rm eff}(T)$ at $T=T^*$, $\beta_{\rm e} = \beta_{\rm e}^{\rm eff}(T^*)$.

The electric exponent $\delta_{\rm e}$ is defined by the singular
dependence (\ref{Ecrit}) of the polarization $P$ at $T=T_c$, under
weak electric field $E$.
Defining the effective exponent $\delta_{\rm e}^{\rm eff}$ as
\begin{equation} \label{eq:delta0fp}
\delta^{\rm eff}(E) = \left( \frac{\partial\ln P}{\partial\ln E} \right)^{-1}.
\end{equation}
and denoting the minimum point of the plot $\delta^{\rm eff}(E)$ as $E^*$, 
we have $\delta_{\rm e} = \delta_{\rm e}^{\rm eff}(E^*)$.

Using the von Nemann entropy at $T=T_c$, we define the effective electric
central charge as
\begin{equation}
c_{\rm e}^{\rm eff}(L) = 6 \frac{\partial S^{\rm e}_{\rm N}}{\partial \ln L}
\end{equation}
and $c_{\rm e}=\lim_{L\to\infty} c_{\rm e}^{\rm eff}(L)$.

\renewcommand{\theequation}{4.\arabic{equation}}
\setcounter{equation}{0}

\section{Numerical analysis of magnetic exponents} \label{Sec.4}
In all considered cases, for simplicity we fix the vertex energy 
$\epsilon_a=0$, i.e., $a=a_c=1$.
The value of the critical temperature $T_c$ is set to 1.
In what follows, the truncation parameter $D=1000$ in all $L$-dependent plots,
while $D=200$ in all dependences of the effective exponents on the deviation 
from the critical temperature $\Delta T = T_c-T$ or the applied magnetic 
(electric) $H$ ($E$) field.
The critical hypersurface (\ref{crit}) of the ferroelectric-A phase 
is considered.

\begin{figure}
\centering
\includegraphics[width=0.45\textwidth,clip]{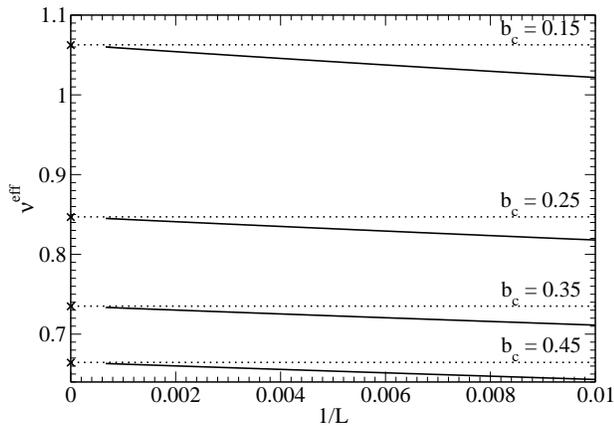}
\caption{The symmetric eight-vertex model with $c=d$: the dependence
of the effective critical index $\nu^{\rm eff}$ on the inverse system size $1/L$,
for four values of the critical vertex weight $b_c = 0,15,0.25,0.35$ and $0.45$.
As $1/L$ goes to $0$, the linear $a+b/L$ fittings of $\nu^{\rm eff}(L)$ give
the asymptotic $\nu$-values denoted by crosses.  
Baxter's exact values are represented by dotted lines.}
\label{obr:nu1}
\end{figure}

\begin{figure}
\centering
\includegraphics[width=0.45\textwidth,clip]{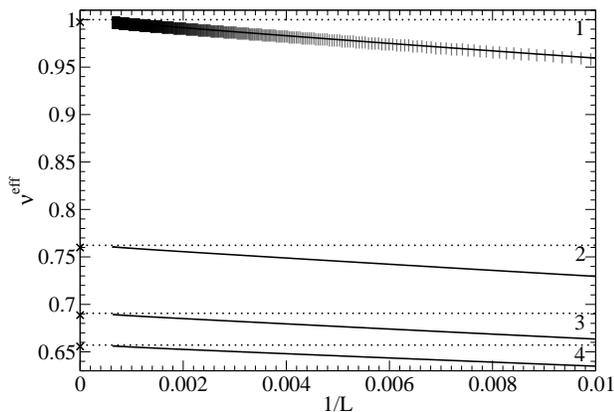}
\caption{The symmetric eight-vertex model with $c\ne d$: the dependence
of $\nu^{\rm eff}$ on $1/L$, for four choices of the critical vertex weights 
(\ref{cases}).
The asymptotic $\nu$-values are denoted by crosses, Baxter's exact values 
are represented by dotted lines.}
\label{obr:nu2}
\end{figure}

The symmetric eight-vertex model with $c=d$ is then defined by
\begin{equation}
b_c , \qquad c_c = \frac{1-b_c}{2} .
\end{equation}
The values of $b_c$ under consideration are $0.15,0.25,0.35$ and $0.45$.

In the case of the symmetric eight-vertex model with $c\ne d$, we consider
four choices of vertex weights:
\begin{equation} \label{cases}
\begin{array}{llll}
1: & b_c = 0.15 & c_c=0.60 & d_c=0.25 \cr
2: & b_c = 0.25 & c_c=0.15 & d_c=0.60 \cr
3: & b_c = 0.35 & c_c=0.50 & d_c=0.15 \cr
4: & b_c = 0.45 & c_c=0.20 & d_c=0.35
\end{array}
\end{equation}

In this section, our numerical method is first tested within the framework of 
the magnetic formulation, with the known Baxter's values of critical exponents.
The obtained numerical results will be first presented in figures to document
visually their accuracy, then the numerical values obtained via asymptotic 
fits will be tabulated and compared with the exact values in Tab. I which is
situated at the end of this section.

The effective exponent $\nu^{\rm eff}$ as a function of the inverse system size 
$1/L$ is pictured in Fig. \ref{obr:nu1} for $c=d$ and in Fig. \ref{obr:nu2} 
for $c\ne d$.
As $1/L$ goes to $0$, the linear $a+b/L$ fittings of $\nu^{\rm eff}(L)$ give
the asymptotic $\nu$-values denoted by crosses which are close to the Baxter's 
exact values of $\nu$ represented by the horizontal dotted lines.
The number of individual values of $L$ used in the numerical calculation
is documented by vertical segments in Fig. \ref{obr:nu2} on the plot 
corresponding to the choice 1 of vertex weights; we recall that the difference
between segments corresponds to the increase of $L$ by 2.
We see that as $1/L\to 0$ the point set is quasi-continuous. 
Since $\ln(L+2)-\ln L \sim 2/L$ for large $L$, the discrete evaluation of 
the derivative with respect to $\ln L$ is accurate.

\begin{figure}
\centering
\includegraphics[width=0.45\textwidth,clip]{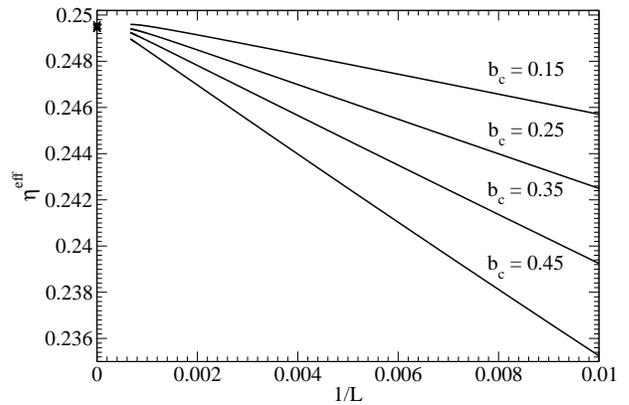}
\caption{The symmetric eight-vertex model with $c=d$: the dependence of 
the effective critical index $\eta^{\rm eff}$ on the inverse system size $1/L$, 
for four values of the critical vertex weight $b_c = 0,15,0.25,0.35$ and $0.45$.
In all cases, as $1/L\to 0$ $\eta^{\rm eff}$ goes to $1/4$.}
\label{obr:eta1}
\end{figure}

\begin{figure}
\centering
\includegraphics[width=0.45\textwidth,clip]{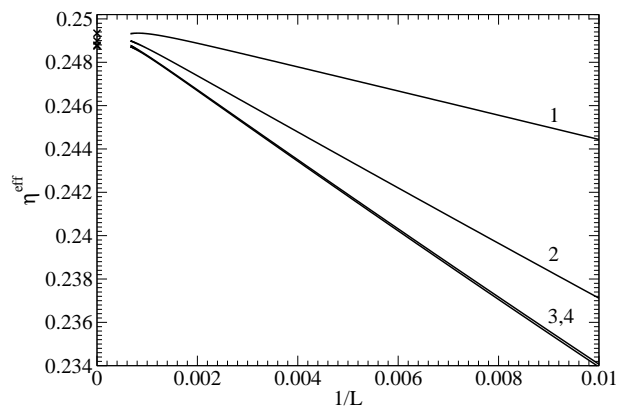}
\caption{The symmetric eight-vertex model with $c\ne d$: the dependence of
$\eta^{\rm eff}$ on $1/L$, for four choices of the critical vertex weights 
(\ref{cases}).
In all cases, $\eta^{\rm eff}$ goes asymptotically to $1/4$.}
\label{obr:eta2}
\end{figure}

The dependence of the effective exponent $\eta^{\rm eff}$ on the inverse system 
size $1/L$ is presented in Fig. \ref{obr:eta1} for $c=d$ and in 
Fig. \ref{obr:eta2} for $c\ne d$.
As $L$ increases, all curves converge to the Ising value $\eta=1/4$ as it
should be for a weakly universal critical theory.
Note that the curves corresponding to choices 3 and 4 in Fig. \ref{obr:eta2} 
are indistinguishable in the present zoom. 

\begin{figure}
\centering
\includegraphics[width=0.45\textwidth,clip]{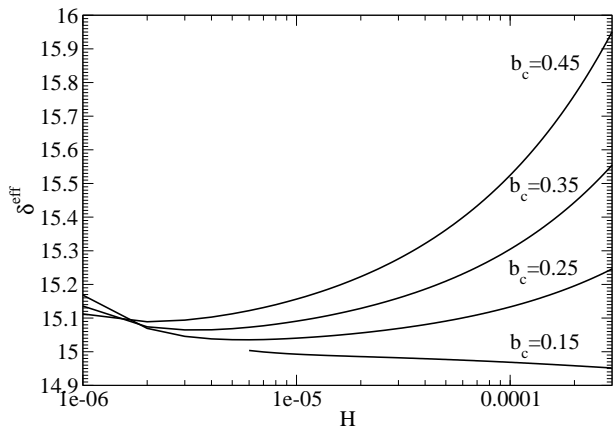}
\caption{The symmetric eight-vertex model with $c=d$: the dependence
of the effective critical exponent $\delta^{\rm eff}$ on the applied magnetic 
field $H$, for four values of the critical vertex weight $b_c = 0,15,0.25,0.35$
and $0.45$.
The actual $\delta$-values, identified with the minimum points of the plots,
are close to the exact Baxter's result $\delta=15$.}
\label{obr:delta1}
\end{figure}

\begin{figure}
\centering
\includegraphics[width=0.45\textwidth,clip]{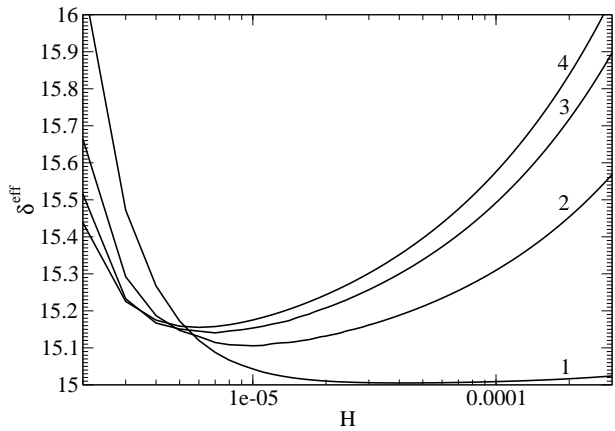}
\caption{The symmetric eight-vertex model with $c\ne d$: the dependence of
$\delta^{\rm eff}$ on the magnetic field $H$, for four choices of the critical 
vertex weights (\ref{cases}).}
\label{obr:delta2}
\end{figure}

In the logarithmic scale, the plots of the effective exponent $\delta^{\rm eff}$ 
versus the applied magnetic field $H$ are presented in Fig. \ref{obr:delta1} 
for $c=d$ and in Fig. \ref{obr:delta2} for $c\ne d$.
The actual $\delta$-values are identified with the minimum points of the plots.
They are close to the Baxter's constant prediction $\delta=15$.
The only exception is the plot for $b_c=0.15$ which does not exhibit a minimum
and therefore we fitted the original dependence (\ref{Hcrit}).

\begin{figure}
\centering
\includegraphics[width=0.45\textwidth,clip]{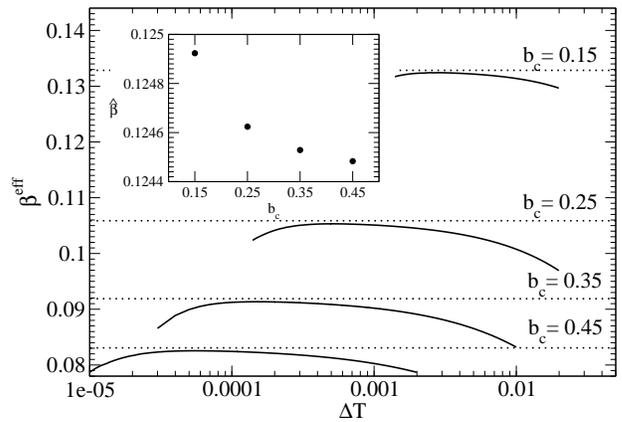}
\caption{The symmetric eight-vertex model with $c=d$: the effective
exponent $\beta^{\rm eff}$ versus the deviation from the critical point
$\Delta T = T_c-T$, for four values of the critical vertex weight 
$b_c = 0,15,0.25,0.35$ and $0.45$. 
The $\beta$-values are identified with the maximum points of the plots,
Baxter's exact values are represented by dotted lines.
The inset documents an almost constant dependence of the rescaled exponent
$\hat{\beta}=\beta/\nu\sim 1/8$ on $b_c$.}
\label{obr:beta1}
\end{figure}

\begin{figure}
\centering
\includegraphics[width=0.45\textwidth,clip]{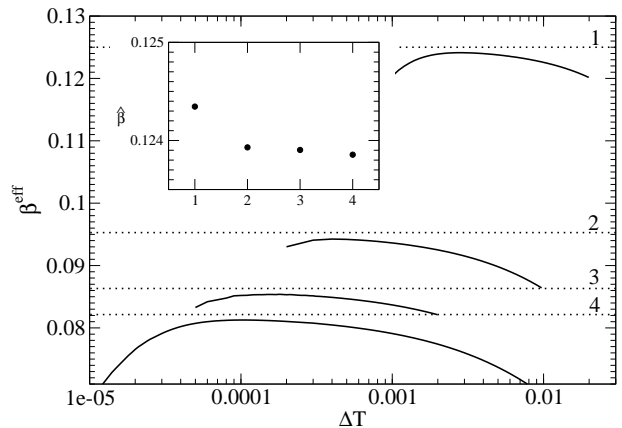}
\caption{The symmetric eight-vertex model with $c\ne d$: the dependence of 
$\beta^{\rm eff}$ on $\Delta T$ for four choices of the critical vertex 
weights (\ref{cases}). 
Baxter's exact values are represented by dotted lines.
The inset brings the dependence of $\hat{\beta}$ on $b_c$.}
\label{obr:beta2}
\end{figure}

In the logarithmic scale, the numerical results for the effective exponent 
$\beta^{\rm eff}$ as the function of the deviation from the critical temperature 
$\Delta T$ are presented in Fig. \ref{obr:beta1} for $c=d$ and in 
Fig. \ref{obr:beta2} for $c\ne d$.
The plots of $\beta^{\rm eff}(\Delta T)$ exhibit maxima values close to the 
Baxter exact results for $\beta$ (dotted lines) as it should be.
The insets of the figures show the model's dependence of the exponent ratio
$\hat{\beta}=\beta/\nu$.
In spite of a slight dispersion of the results, $\hat{\beta}$ is close to 
the Ising value $\beta_{\rm I}=1/8$, in agreement with the concept
of weak universality.

\begin{figure}
\centering
\includegraphics[width=0.45\textwidth,clip]{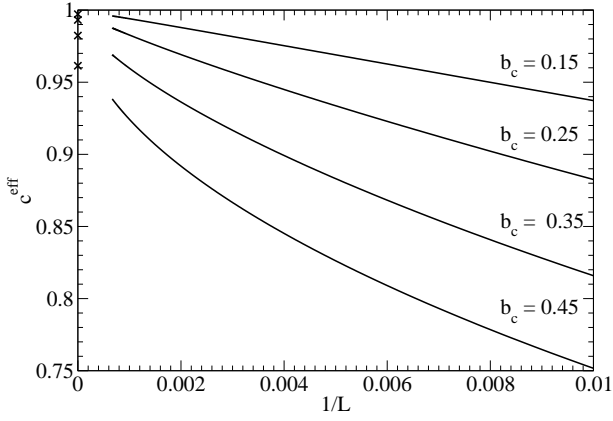}
\caption{The symmetric eight-vertex model with $c=d$: the dependence of 
the magnetic effective central charge $c^{\rm eff}$ on $1/L$, for four values of 
the critical vertex weight $b_c = 0,15,0.25,0.35$ and $0.45$. 
As $1/L\to 0$, all curves tend to the central charge $c=1$.}
\label{obr:c}
\end{figure}

For the vertex weights $c=d$, the magnetic effective central charge 
$c^{\rm eff}$ is presented as a function of the inverse system size $1/L$ 
in Fig. \ref{obr:c}.
For all four values of the critical vertex weight $b_c = 0,15,0.25,0.35$ and 
$0.45$, as $1/L\to 0$ the plots tend to the value $c=1$ which is the central
charge of the weakly universal symmetric eight-vertex model \cite{Baxterbook}. 

\begin{table} \label{Table1}
\begin{tabular}{||l|l|l|l||} \hline
Fig. \ref{obr:nu1} & $b_c=0.15$ & $\nu^{\rm num}=1.0624$ & 
$\nu^{\rm exact}=1.0628$ \\ \cline{2-4}
& $\phantom{b_c}=0.25$ & $\phantom{\nu^{\rm num}}=0.8468$ & 
$\phantom{\nu^{\rm exact}}=0.8470$ \\ \cline{2-4}
& $\phantom{b_c}=0.35$ & $\phantom{\nu^{\rm num}}=0.7349$ & 
$\phantom{\nu^{\rm exact}}=0.7351$ \\ \cline{2-4}  
& $\phantom{b_c}=0.45$ & $\phantom{\nu^{\rm num}}=0.6643$ & 
$\phantom{\nu^{\rm exact}}=0.6646$ \\ \hline
Fig. \ref{obr:nu2} & 1 & $\nu^{\rm num}=0.9975$ & 
$\nu^{\rm exact}=1.0000$ \\ \cline{2-4}
& 2 & $\phantom{\nu^{\rm num}}=0.7599$ & $\phantom{\nu^{\rm exact}}=0.7622$ 
\\ \cline{2-4}
& 3 & $\phantom{\nu^{\rm num}}=0.6887$ & $\phantom{\nu^{\rm exact}}=0.6906$ 
\\ \cline{2-4}  
& 4 & $\phantom{\nu^{\rm num}}=0.6557$ & $\phantom{\nu^{\rm exact}}=0.6572$ 
\\ \hline
Fig. \ref{obr:eta1} & $b_c=0.15$ & $\eta^{\rm num}=0.2496$ & 
$\eta^{\rm exact}=1/4$ \\ \cline{2-4}
& $\phantom{b_c}=0.25$ & $\phantom{\eta^{\rm num}}=0.2494$ & 
$\phantom{\eta^{\rm exact}}=1/4$ \\ \cline{2-4}
& $\phantom{b_c}=0.35$ & $\phantom{\eta^{\rm num}}=0.2495$ & 
$\phantom{\eta^{\rm exact}}=1/4$ \\ \cline{2-4}  
& $\phantom{b_c}=0.45$ & $\phantom{\eta^{\rm num}}=0.2494$ & 
$\phantom{\eta^{\rm exact}}=1/4$ \\ \hline
Fig. \ref{obr:eta2} & 1 & $\eta^{\rm num}=0.2493$ & 
$\eta^{\rm exact}=1/4$ \\ \cline{2-4}
& 2 & $\phantom{\eta^{\rm num}}=0.2490$ & $\phantom{\eta^{\rm exact}}=1/4$ 
\\ \cline{2-4}
& 3 & $\phantom{\eta^{\rm num}}=0.2488$ & $\phantom{\eta^{\rm exact}}=1/4$ 
\\ \cline{2-4}  
& 4 & $\phantom{\eta^{\rm num}}=0.2487$ & $\phantom{\eta^{\rm exact}}=1/4$ 
\\ \hline
Fig. \ref{obr:delta1} & $b_c=0.15$ & $\delta^{\rm num}=15.0355$ & 
$\delta^{\rm exact}=15$ \\ \cline{2-4}
& $\phantom{b_c}=0.25$ & $\phantom{\delta^{\rm num}}=15.0648$ & 
$\phantom{\delta^{\rm exact}}=15$ \\ \cline{2-4}
& $\phantom{b_c}=0.35$ & $\phantom{\delta^{\rm num}}=15.0891$ & 
$\phantom{\delta^{\rm exact}}=15$ \\ \cline{2-4}  
& $\phantom{b_c}=0.45$ & $\phantom{\delta^{\rm num}}=14.9811$ & 
$\phantom{\delta^{\rm exact}}=15$ \\ \hline
Fig. \ref{obr:delta2} & 1 & $\delta^{\rm num}=15.0051$ & 
$\delta^{\rm exact}=15$ \\ \cline{2-4}
& 2 & $\phantom{\delta^{\rm num}}=15.1055$ & $\phantom{\delta^{\rm exact}}=15$ 
\\ \cline{2-4}
& 3 & $\phantom{\delta^{\rm num}}=15.1404$ & $\phantom{\delta^{\rm exact}}=15$ 
\\ \cline{2-4}  
& 4 & $\phantom{\delta^{\rm num}}=15.1554$ & $\phantom{\delta^{\rm exact}}=15$ 
\\ \hline
Fig. \ref{obr:beta1} & $b_c=0.15$ & $\beta^{\rm num}=0.1324$ & 
$\beta^{\rm exact}=0.1328$ \\ \cline{2-4}
& $\phantom{b_c}=0.25$ & $\phantom{\beta^{\rm num}}=0.1053$ & 
$\phantom{\beta^{\rm exact}}=0.1059$ \\ \cline{2-4}
& $\phantom{b_c}=0.35$ & $\phantom{\beta^{\rm num}}=0.0913$ & 
$\phantom{\beta^{\rm exact}}=0.0919$ \\ \cline{2-4}  
& $\phantom{b_c}=0.45$ & $\phantom{\beta^{\rm num}}=0.0826$ & 
$\phantom{\beta^{\rm exact}}=0.0831$ \\ \hline
Fig. \ref{obr:beta2} & 1 & $\beta^{\rm num}=0.1241$ & 
$\beta^{\rm exact}=0.1250$ \\ \cline{2-4}
& 2 & $\phantom{\beta^{\rm num}}=0.0942$ & $\phantom{\beta^{\rm exact}}=0.0953$ 
\\ \cline{2-4}
& 3 & $\phantom{\beta^{\rm num}}=0.0854$ & $\phantom{\beta^{\rm exact}}=0.0863$ 
\\ \cline{2-4}  
& 4 & $\phantom{\beta^{\rm num}}=0.0813$ & $\phantom{\beta^{\rm exact}}=0.0821$ 
\\ \hline
Fig. \ref{obr:c} & $b_c=0.15$ & $c^{\rm num}=0.9972$ & 
$c^{\rm exact}=1$ \\ \cline{2-4}
& $\phantom{b_c}=0.25$ & $\phantom{c^{\rm num}}=0.9932$ & 
$\phantom{c^{\rm exact}}=1$ \\ \cline{2-4}
& $\phantom{b_c}=0.35$ & $\phantom{c^{\rm num}}=0.9823$ & 
$\phantom{c^{\rm exact}}=1$ \\ \cline{2-4}  
& $\phantom{b_c}=0.45$ & $\phantom{c^{\rm num}}=0.9614$ & 
$\phantom{c^{\rm exact}}=1$ \\ \hline
\end{tabular}
\caption{Numerical data for the magnetic exponents and central charge obtained 
from Figs. \ref{obr:nu1}-\ref{obr:c} and compared with the Baxter's exact 
results.}
\end{table}

Numerical data for the magnetic exponents obtained from 
Figs. \ref{obr:nu1}-\ref{obr:c} are tabulated in Tab. I. 
The comparison with Baxter's exact results confirms a high precision
of our numerical results.

\renewcommand{\theequation}{5.\arabic{equation}}
\setcounter{equation}{0}

\section{Numerical analysis of electric exponents} \label{Sec.5}
\begin{figure}
\centering
\includegraphics[width=0.45\textwidth,clip]{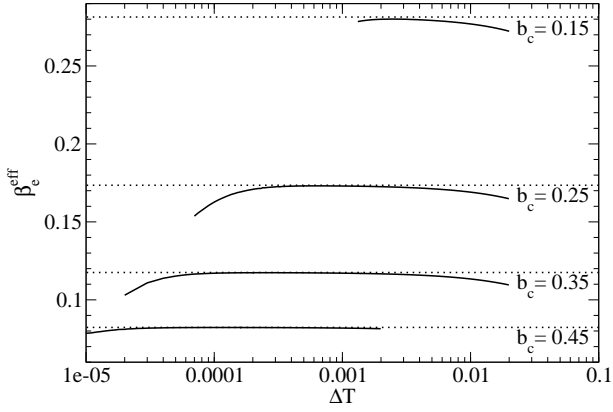}
\caption{The symmetric eight-vertex model with $c=d$: the electric effective  
exponent $\beta_{\rm e}^{\rm eff}$ versus the deviation from the critical point
$\Delta T = T_c-T$, for four values of the critical vertex weight 
$b_c = 0,15,0.25,0.35$ and $0.45$. 
The $\beta_{\rm e}$-values are identified with the maximum points of the plots,
Baxter's exact values are represented by dotted lines.}
\label{obr:betae1}
\end{figure}

\begin{figure}
\centering
\includegraphics[width=0.45\textwidth,clip]{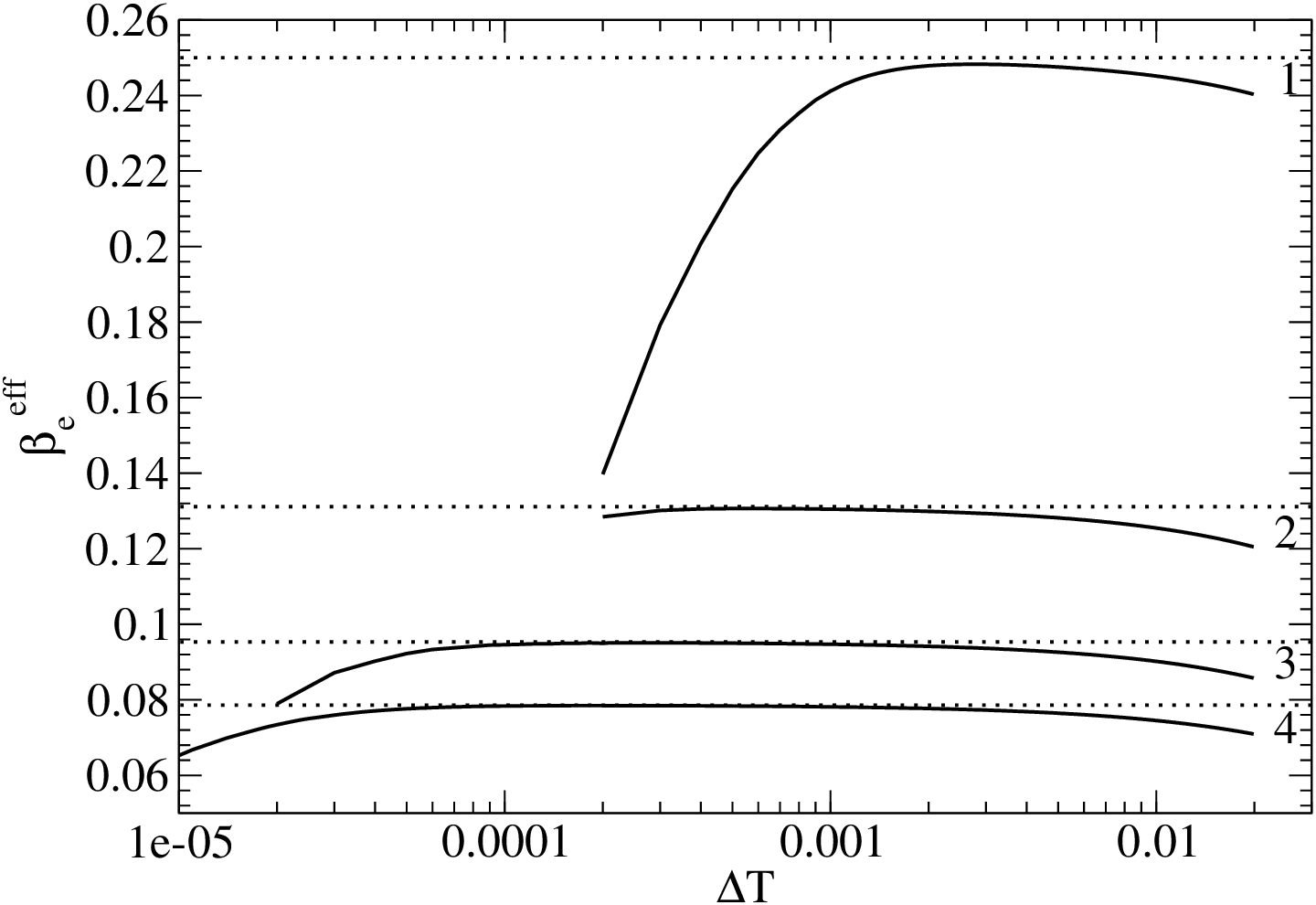}
\caption{The symmetric eight-vertex model with $c\ne d$: the dependence of 
$\beta_{\rm e}^{\rm eff}$ on $\Delta T$, for four choices of the critical vertex 
weights (\ref{cases}). 
Baxter's exact values are represented by dotted lines.}
\label{obr:betae2}
\end{figure}

In the logarithmic scale, the numerical results for the effective exponent 
$\beta_{\rm e}^{\rm eff}$ as the function of the deviation from the critical 
temperature $\Delta T$ are presented in Fig. \ref{obr:betae1} for $c=d$ and 
in Fig. \ref{obr:betae2} for $c\ne d$.
The plots of $\beta^{\rm eff}(\Delta T)$ exhibit maxima close to the Baxter 
exact result for $\beta_{\rm e}$ (\ref{betae}) (horizontal dotted lines).
This fact confirms the adequacy of our numerical results also
in the electric format.

Let us combine Baxter's exact result for the electric exponent 
$\beta_{\rm e}$ (\ref{betae}) with the equality between magnetic and 
electric $\nu$-indices (\ref{nue}) and the scaling relations (\ref{scalinge}). 
The exponents $\eta_{\rm e}$ and $\delta_{\rm e}$ are then given by 
\begin{equation} \label{suggestetae}
\eta_{\rm e} = 1 - \frac{\mu}{\pi} , \qquad 
\delta_{\rm e} = \frac{3\pi+\mu}{\pi-\mu} .
\end{equation}
As was explained before, the dependence of the electric exponents $\eta_{\rm e}$ 
and $\delta_{\rm e}$, defined just at the critical point, on model's parameters 
means that the original electric version of the symmetric eight-vertex model 
cannot be weakly universal, but it is fully non-universal. 

\begin{figure}
\centering
\includegraphics[width=0.45\textwidth,clip]{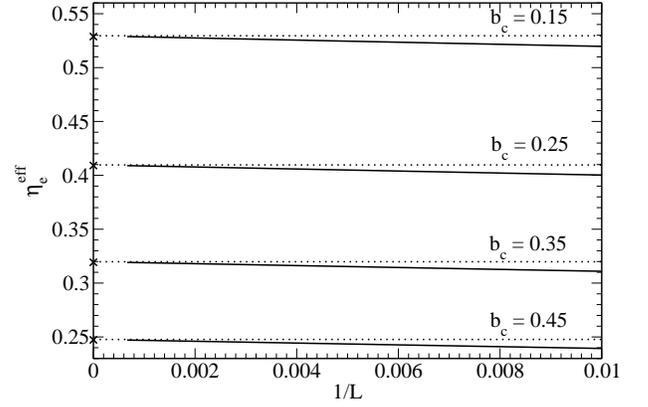}
\caption{The symmetric eight-vertex model with $c=d$: the dependence
of the effective critical exponent $\eta_{\rm e}^{\rm eff}$ on 
the inverse system size $1/L$, for four values of the critical vertex weight 
$b_c = 0,15,0.25,0.35$ and $0.45$.
The suggested values obtained from (\ref{suggestetae}) are represented 
by dotted lines.}
\label{obr:etae1}
\end{figure}

\begin{figure}
\centering
\includegraphics[width=0.45\textwidth,clip]{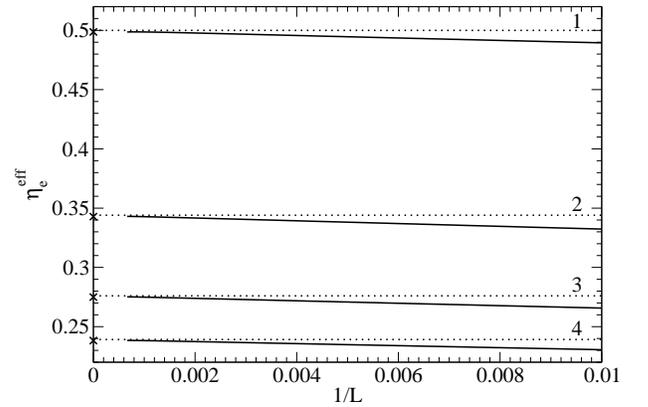}
\caption{The symmetric eight-vertex model with $c\ne d$: the dependence of
$\eta_{\rm e}^{\rm eff}$ on $1/L$, for four choices of the critical vertex 
weights (\ref{cases}).
The suggested values obtained from (\ref{suggestetae}) are represented 
by dotted lines.}
\label{obr:etae2}
\end{figure}

The plot of the effective exponent $\eta_{\rm e}^{\rm eff}$ versus 
the inverse system size $1/L$ is pictured in Fig. \ref{obr:etae1} for $c=d$ 
and in Fig. \ref{obr:etae2} for $c\ne d$.
As $L$ increases, the curves converge to the asymptotic values (crosses) which 
are in agreement with our suggested formula (\ref{suggestetae}). 
 
\begin{figure}
\centering
\includegraphics[width=0.45\textwidth,clip]{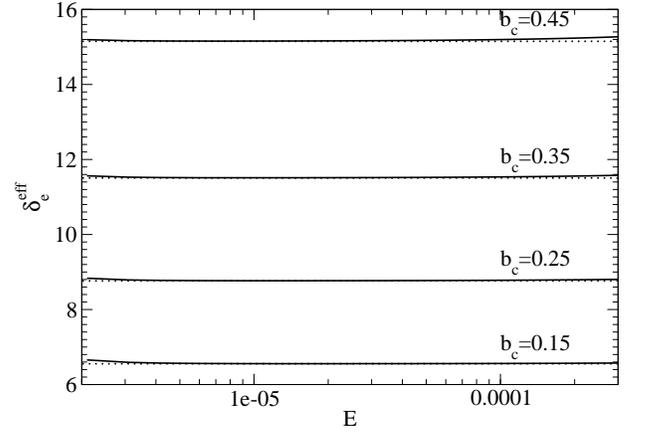}
\caption{The symmetric eight-vertex model with $c=d$: the dependence
of the effective critical exponent $\delta_{\rm e}^{\rm eff}$ on the 
the applied electric field $E$, for four values of the critical vertex weight 
$b_c = 0,15,0.25,0.35$ and $0.45$.
The $\delta_{\rm e}$-values are identified with the minimum points of the plots,
the suggested values (\ref{suggestetae}) are represented by dotted lines.}
\label{obr:deltae1}
\end{figure}

\begin{figure}
\centering
\includegraphics[width=0.45\textwidth,clip]{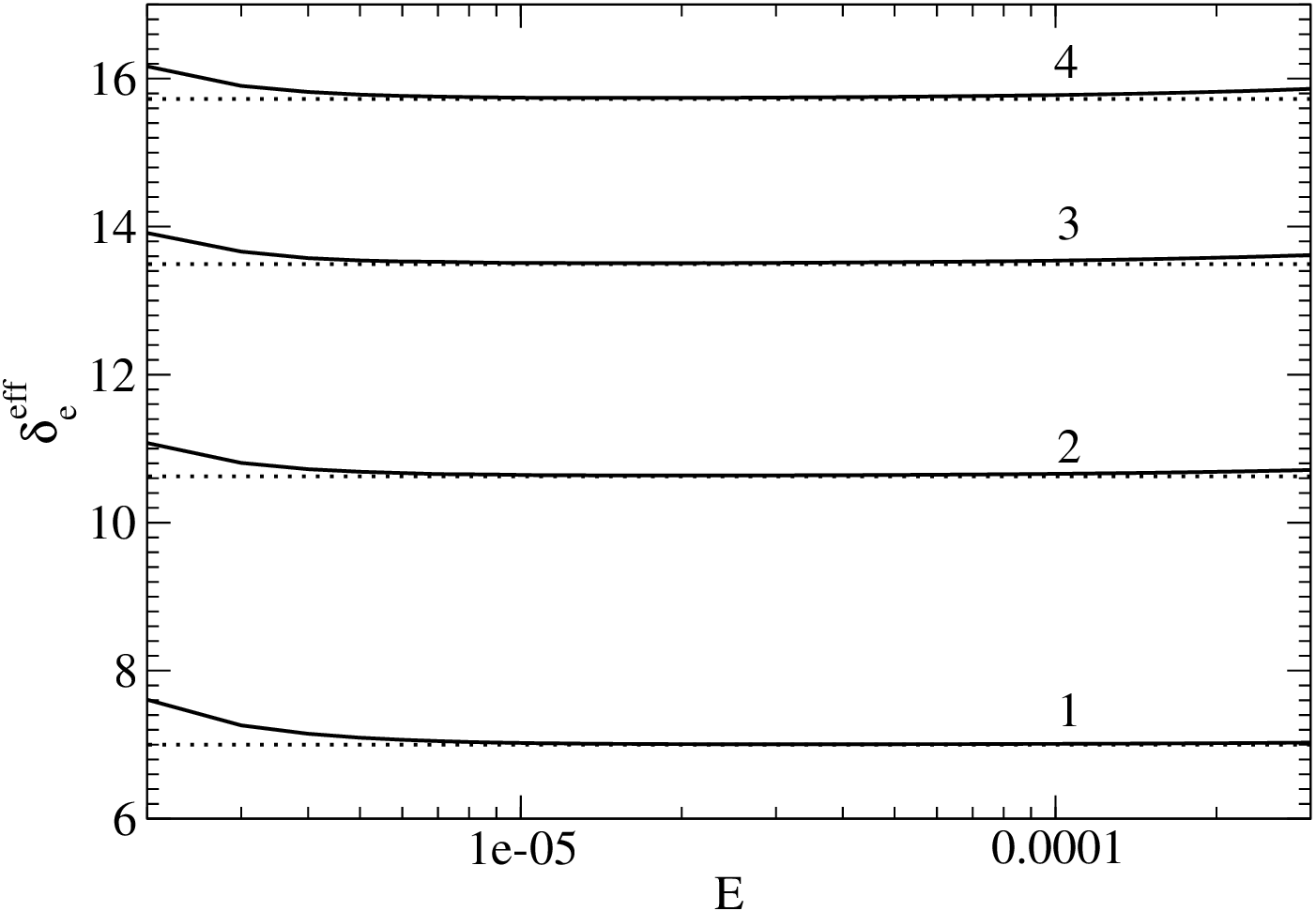}
\caption{The symmetric eight-vertex model with $c\ne d$: the dependence of
$\delta_{\rm e}^{\rm eff}$ on the electric field $E$, for four choices of 
the critical vertex weights (\ref{cases}).
The suggested values (\ref{suggestetae}) are represented by dotted 
lines.}
\label{obr:deltae2}
\end{figure}

In the logarithmic scale, the plots of the effective exponent 
$\delta_{\rm e}^{\rm eff}$ versus the applied electric field $E$ are pictured in 
Fig. \ref{obr:deltae1} for $c=d$ and in Fig. \ref{obr:deltae2} for $c\ne d$.
Note a shallowness of the plots.
The $\delta_{\rm e}$-values are identified with the minimum points of the plots,
the suggested values (\ref{suggestetae}) are represented by dotted lines.

\begin{figure}
\centering
\includegraphics[width=0.4\textwidth,clip]{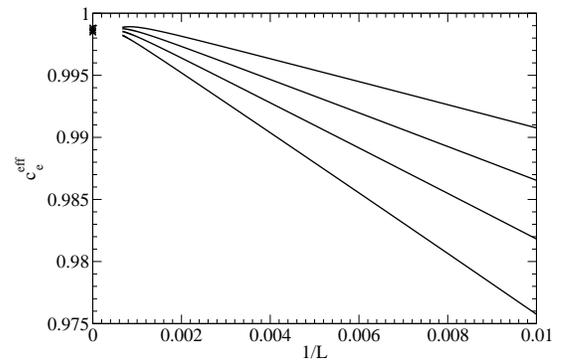}
\caption{The symmetric eight-vertex model with $c=d$: the dependence of 
the effective electric central charge $c_{\rm e}^{\rm eff}$ on $1/L$, for four 
values of the critical vertex weight $b_c = 0,15,0.25,0.35$ and $0.45$. 
As $1/L\to 0$, all curves tend to $c_{\rm e}=1$.}
\label{obr:ce}
\end{figure}

For the vertex weights $c=d$, the dependence of the electric effective 
central charge $c_{\rm e}^{\rm eff}$ on the inverse system size $1/L$ 
is pictured in Fig. \ref{obr:ce}.
As before for the magnetic case, for all four values of the critical vertex 
weight $b_c = 0,15,0.25,0.35$ and $0.45$, the plots tend as $1/L\to 0$
to the same value $c_{\rm e}=1$. 

\begin{table} \label{Table2}
\begin{tabular}{||l|l|l|l||} \hline
Fig. \ref{obr:betae1} & $b_c=0.15$ & $\beta_{\rm e}^{\rm num}=0.2801$ & 
$\beta_{\rm e}^{\rm exact}=0.2814$ \\ \cline{2-4}
& $\phantom{b_c}=0.25$ & $\phantom{\beta_{\rm e}^{\rm num}}=0.1731$ & 
$\phantom{\beta_{\rm e}^{\rm exact}}=0.1735$ \\ \cline{2-4}
& $\phantom{b_c}=0.35$ & $\phantom{\beta_{\rm e}^{\rm num}}=0.1174$ & 
$\phantom{\beta_{\rm e}^{\rm exact}}=0.1175$ \\ \cline{2-4}  
& $\phantom{b_c}=0.45$ & $\phantom{\beta_{\rm e}^{\rm num}}=0.0822$ & 
$\phantom{\beta_{\rm e}^{\rm exact}}=0.0823$ \\ \hline
Fig. \ref{obr:betae2} & 1 & $\beta_{\rm e}^{\rm num}=0.2483$ & 
$\beta_{\rm e}^{\rm exact}=0.2500$ \\ \cline{2-4}
& 2 & $\phantom{\beta_{\rm e}^{\rm num}}=0.1306$ & 
$\phantom{\beta_{\rm e}^{\rm exact}}=0.1311$ \\ \cline{2-4}
& 3 & $\phantom{\beta_{\rm e}^{\rm num}}=0.0951$ & 
$\phantom{\beta_{\rm e}^{\rm exact}}=0.0953$ \\ \cline{2-4}  
& 4 & $\phantom{\beta_{\rm e}^{\rm num}}=0.0785$ & 
$\phantom{\beta_{\rm e}^{\rm exact}}=0.0786$ 
\\ \hline
Fig. \ref{obr:etae1} & $b_c=0.15$ & $\eta_{\rm e}^{\rm num}=0.5288$ & 
$\eta_{\rm e}^{\rm sugg}=0.5296$ \\ \cline{2-4}
& $\phantom{b_c}=0.25$ & $\phantom{\eta_{\rm e}^{\rm num}}=0.4091$ & 
$\phantom{\eta_{\rm e}^{\rm sugg}}=0.4097$ \\ \cline{2-4}
& $\phantom{b_c}=0.35$ & $\phantom{\eta_{\rm e}^{\rm num}}=0.3193$ & 
$\phantom{\eta_{\rm e}^{\rm sugg}}=0.3198$ \\ \cline{2-4}  
& $\phantom{b_c}=0.45$ & $\phantom{\eta_{\rm e}^{\rm num}}=0.2473$ & 
$\phantom{\eta_{\rm e}^{\rm sugg}}=0.2477$ \\ \hline
Fig. \ref{obr:etae2} & 1 & $\eta_{\rm e}^{\rm num}=0.4987$ & 
$\eta_{\rm e}^{\rm sugg}=0.5000$ \\ \cline{2-4}
& 2 & $\phantom{\eta_{\rm e}^{\rm num}}=0.3428$ & 
$\phantom{\eta_{\rm e}^{\rm sugg}}=0.3440$ \\ \cline{2-4}
& 3 & $\phantom{\eta_{\rm e}^{\rm num}}=0.2750$ & 
$\phantom{\eta_{\rm e}^{\rm sugg}}=0.2760$ \\ \cline{2-4}  
& 4 & $\phantom{\eta_{\rm e}^{\rm num}}=0.2383$ & 
$\phantom{\eta_{\rm e}^{\rm sugg}}=0.2392$ 
\\ \hline
Fig. \ref{obr:deltae1} & $b_c=0.15$ & $\delta_{\rm e}^{\rm num}=6.5561$ & 
$\delta_{\rm e}^{\rm sugg}=6.5539$ \\ \cline{2-4}
& $\phantom{b_c}=0.25$ & $\phantom{\delta_{\rm e}^{\rm num}}=8.7674$ & 
$\phantom{\delta_{\rm e}^{\rm sugg}}=8.7641$ \\ \cline{2-4}
& $\phantom{b_c}=0.35$ & $\phantom{\delta_{\rm e}^{\rm num}}=11.5123$ & 
$\phantom{\delta_{\rm e}^{\rm sugg}}=11.5077$ \\ \cline{2-4}  
& $\phantom{b_c}=0.45$ & $\phantom{\delta_{\rm e}^{\rm num}}=15.1562$ & 
$\phantom{\delta_{\rm e}^{\rm sugg}}=15.1500$ \\ \hline
Fig. \ref{obr:deltae2} & 1 & $\delta_{\rm e}^{\rm num}=7.0054$ & 
$\delta_{\rm e}^{\rm sugg}=7.0000$ \\ \cline{2-4}
& 2 & $\phantom{\delta_{\rm e}^{\rm num}}=10.6365$ & 
$\phantom{\delta_{\rm e}^{\rm sugg}}=10.6265$ \\ \cline{2-4}
& 3 & $\phantom{\delta_{\rm e}^{\rm num}}=13.5057$ & 
$\phantom{\delta_{\rm e}^{\rm sugg}}=13.4928$ \\ \cline{2-4}  
& 4 & $\phantom{\delta_{\rm e}^{\rm num}}=15.7393$ & 
$\phantom{\delta_{\rm e}^{\rm sugg}}=15.7251$ 
\\ \hline
Fig. \ref{obr:ce} & $b_c=0.15$ & $c_{\rm e}^{\rm num}=0.9988$ & 
$c_{\rm e}^{\rm sugg}=1$ \\ \cline{2-4}
& $\phantom{b_c}=0.25$ & $\phantom{c_{\rm e}^{\rm num}}=0.9987$ & 
$\phantom{c_{\rm e}^{\rm sugg}}=1$ \\ \cline{2-4}
& $\phantom{b_c}=0.35$ & $\phantom{c_{\rm e}^{\rm num}}=0.9985$ & 
$\phantom{c_{\rm e}^{\rm sugg}}=1$ \\ \cline{2-4}  
& $\phantom{b_c}=0.45$ & $\phantom{c_{\rm e}^{\rm num}}=0.9985$ & 
$\phantom{c_{\rm e}^{\rm sugg}}=1$ \\ \hline
\end{tabular}
\caption{Numerical data for the electric exponents and central charge obtained 
from Figs. \ref{obr:betae1}-\ref{obr:ce} and compared with the Baxter's exact 
result for $\beta_{\rm e}$ (\ref{betae}) or the values generated
from our suggested formulas (\ref{suggestetae}).}
\end{table}

Numerical data for the electric exponents obtained from 
Figs. \ref{obr:betae1}-\ref{obr:ce} are tabulated in Tab. II. 
The comparison with the Baxter's exact result for $\beta_{\rm e}$ or 
the values obtained from our suggested formulas in Eq. (\ref{suggestetae}) 
shows a high precision of our numerical results.

\renewcommand{\theequation}{6.\arabic{equation}}
\setcounter{equation}{0}

\section{Conclusion} \label{Sec.6}
Baxter solved the symmetric eight-vertex model on the square lattice
within its magnetic formulation of Ising spins on the dual square lattice
with plaquette interactions.
Some of the magnetic critical exponents depend on model's parameters.
Pointing out a freedom in the definition of deviation from the critical 
point, Suzuki proposed a rescaling of critical indices.
The rescaled indices become constant, namely 2D Ising-like, and this
property is known as weak universality.
Weak universality requires that the exponents $\eta$ and $\delta$, which are 
defined just at the critical point and therefore do not depend on 
the definition of the deviation from the critical point, are constant and 
indeed $\eta=1/4$ and $\delta=15$.
We tested our numerical estimates of critical indices against Baxter's exact 
results (dotted lines) in Figs. \ref{obr:nu1}-\ref{obr:beta2}, see also
numerical data in Tab. I, the agreement is very good.

As concerns the original vertex (electric) formulation, Baxter was able
to derive the explicit formula (\ref{betae}) for the critical exponent 
$\beta_{\rm e}$ related to the spontaneous polarization.
The crucial point of our analysis was the equivalence of the exponents 
$\nu$ and $\nu_{\rm e}$ in Eq. (\ref{nue}). 
Combining this relation with Baxter's exact result for $\beta_{\rm e}$ 
(\ref{betae}) and the scaling relations (\ref{scalinge}), the suggested 
exponents $\eta_{\rm e}$ and $\delta_{\rm e}$ (\ref{suggestetae}) turns out to 
be dependent on model's parameters.
As is seen in Figs. \ref{obr:etae1} and \ref{obr:etae2}, the numerical check
of the suggested formula for $\eta_{\rm e}$ is very good.
The same applies to the numerical checks of the suggested formula for 
$\delta_{\rm e}$, see Figs. \ref{obr:deltae1} and \ref{obr:deltae2}.
Since the critical exponents $\eta_{\rm e}$ and $\delta_{\rm e}$ are defined
just at the critical point and therefore are independent of the definition 
of the deviation from the critical point, their dependence on model's 
parameters means that the electric vertex formulation of the model is fully 
non-universal.  
Consequently, in spite of the equivalence of the partition functions, 
the magnetic and electric versions of the model possess different
critical properties.

We believe that this work will be a motivation for a rigorous derivation 
of the suggested formulas (\ref{suggestetae}), maybe by using the QISM 
machinery.
The full non-universality of statistical models is probably not so
exceptional as is generally believed.

\begin{acknowledgments}
We are grateful to Prof. Ingo Peschel for providing us with information
on electric critical exponents for the six-vertex model and to Dr.
Andrej Gendiar for discussions about numerics related to the CTMRG method.  
This work was supported by the project EXSES APVV-16-0186 and VEGA Grants 
No. 2/0130/15 and No. 2/0015/15.
\end{acknowledgments}

\end{document}